\documentclass{article}

\usepackage{PRIMEarxiv}

\usepackage[utf8]{inputenc} 
\usepackage[T1]{fontenc}    
\usepackage{hyperref}       
\usepackage{url}            
\usepackage{booktabs}       
\usepackage{amsfonts}       
\usepackage{nicefrac}       
\usepackage{microtype}      
\usepackage{lipsum}
\usepackage{fancyhdr}       
\usepackage{graphicx}       
\graphicspath{{media/}}     

\title{Blockchain and Carbon Markets: Standards Overview
\thanks{\textit{\underline{Citation}}: 
\textbf{TODO: Authors. Title. Pages.... DOI:000000/11111.}} 
}

\author{
  Pedro Baiz \\
  Clearly; Imperial College London \\
  \texttt{pedro@clearly.earth; p.m.baiz@imperial.ac.uk} \\
}

\begin{document}
\maketitle

\begin{center}
    \textbf{Review Paper}
\end{center}

\begin{abstract}
The increasing significance of sustainability considerations within both public spheres (such as policies and regulations) and private sectors (including voluntary commitments by major multinational corporations) underscores the imperative to harness cutting-edge technological advancements. This is essential to ensure that the momentum of this trend translates into tangible outcomes, thwarting phenomena like greenwashing and upholding high standards of integrity, all while expediting progress through automation. This paper focuses specifically on carbon markets, which, after enduring years of  confusion and controversy, may finally be on the brink of converging toward internationally recognized minimum standards. Beginning with an introduction to fundamental concepts pertaining to carbon markets and Distributed Ledger Technologies (DLTs), the paper proceeds to dissect the challenges and opportunities within this burgeoning field. Its primary contribution lies in offering a comprehensive overview of recent developments across various initiatives (such as ICVCM, IETA/WorldBank/CAD Trust, IEEE/ISO) and providing a layered analysis of the entire ecosystem. This framework aids in understanding and prioritising future endeavours. Ultimately, the paper furnishes a set of recommendations aimed at bolstering scalability and fostering widespread adoption of best practices within international markets.




\end{abstract}



\section{Introduction}

Recent setbacks to the crypto space (e.g. FTX, Terra) could give the impression that the promised benefits of DLT are imaginary, but that is very far from the truth. A good way to explain what has (and will continue) to happen can be summarised with the computer science term GIGO (garbage in, garbage out). The catastrophic failures seen in 2022 demonstrates that garbage in the blockchain space could come from many different angles, from the actual "digital asset" themselves to the way the entire organisation or ecosystem is "governed". Standards and regulations are key to prevent GIGO and therefore represent one of the main (if not the main) blocker to scalable and sustainable business models. 

The present document aims to provide an overview and review of the DLT space for Carbon Markets. The review covers activities across all areas, from academia to industrial and international non-profit organisations (e.g. World Bank), specially to support widespread market adoption. The review paper aims to particularly address the following challenges of this growing domain:

\begin{itemize}
\item Lack of coordination between key international organisations related to this broad and growing space. For example: IETA, ICVCM, ISO, IEEE, World Bank, among many others (specially with for-profit organisations which due to competition do not always engage others).
\item Lack of clarity on what exactly this space involves (scoping). In order to move forward we need to avoid confusion of terminology, we need to identify the key players \& organisations, describe clearly the obstacles and enablers, etc. 
\end{itemize}

As this document will demonstrate, the number of initiatives focused on this space has ballooned in recent years, mainly for 2 reasons: 1) Growing demand in carbon markets (e.g. voluntary, regulated, article 6 Paris Agreement); and 2) Realisation from the DLT community that "digital assets" with a direct link to valuable assets (e.g. commodities, physical assets like buildings) have a better chance of success. Unfortunately, multiple DLT projects working with carbon have demonstrated a lack of proper understanding of the complexities of carbon markets, and vice-versa, carbon market experts lack a proper understanding of DLTs, e.g. \cite{TIME2022}. This paper aims to provide a comprehensive overview that enables experts from any domain to gather a good grasp of the current state-of-the-art and more importantly where we are heading. 

\section{Fundamentals of Carbon Markets and DLTs}

\begin{figure}
  \includegraphics[width=\linewidth]{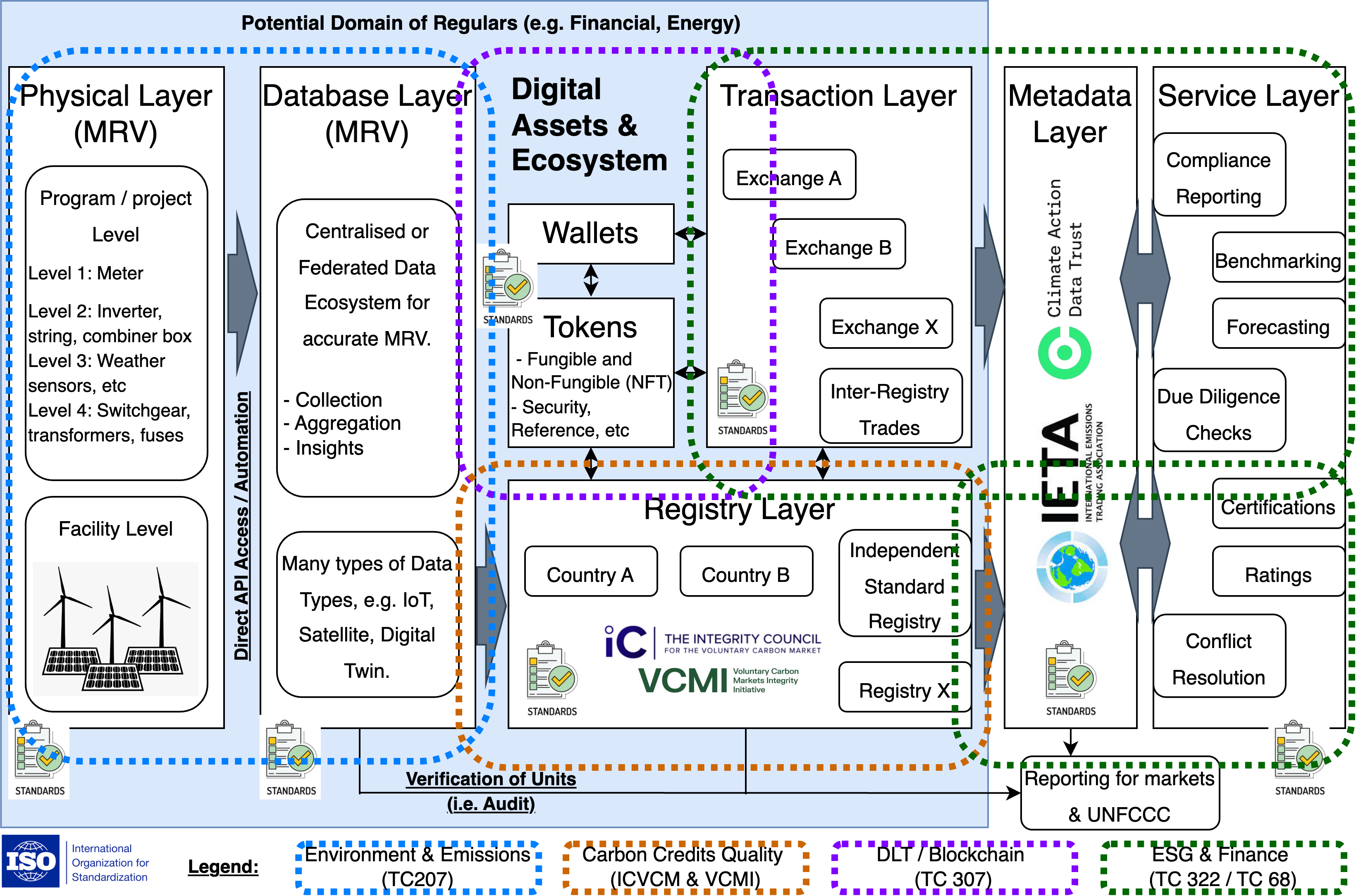}
  \caption{End-to-End Ecosystem of DLT Carbon Markets.}
  \label{fig:EndToEndEcosystem}
\end{figure}

As this paper is meant to be used by many different types of experts, it is important to provide basic definitions that bring experts from different domains up to the same basic level of understanding (e.g. technology vs sustainability). This section will introduce basic concepts that will be referred to in other sections of this document. Figure \ref{fig:EndToEndEcosystem} provides an overview of the entire ecosystem. The figure of the ecosystem shows how it can be broken into different areas that relate to the physical layer (actual emissions measurements) or financial markets (e.g. exchanges). By using the concept of layers, such as transaction, registry, metadata and service, etc, Figure \ref{fig:EndToEndEcosystem} breaks the ecosystem into distinct areas, each of which requires its own type of expertise (it is in fact extremely difficult to find experts that understand all these areas in detail). In the following subsections, a deeper dive will be provided for the broad area of Carbon Markets and DLTs.

\subsection{Carbon Credits/Offset/Market}

Carbon markets are mechanisms that allow countries, companies, and other organisations to buy and sell credits for greenhouse gas emissions. The goal of carbon markets is to create financial incentives for reducing greenhouse gas emissions. Credits can be generated from either: 1) Avoidance/reduction or 2) Removal/sequestration. 

There are two main types of carbon markets:
\begin{itemize}
  \item Regulated / Cap-and-trade systems: This system is imposed by a government or regulatory body which sets a cap on the total amount of greenhouse gases that can be emitted within a specific time period. Companies or organisations that need to emit greenhouse gases must purchase allowances or credits from those who have reduced their emissions below the cap. The overall cap is gradually reduced over time, encouraging companies to find ways to reduce their emissions. Currently we have this type of market in multiple jurisdictions such as California, EU or China. The most polluting economic sectors are initially covered (e.g. energy, cement, steel) and with time others should eventually follow (e.g. Germany just started to cover building and transport sectors).
  \item Voluntary / Carbon offset programs: Carbon offset programs allow companies or individuals to offset their own greenhouse gas emissions by purchasing credits from projects that reduce or remove greenhouse gases from the atmosphere. These projects can include renewable energy projects, reforestation projects, methane capture projects, among many others. As this market is unregulated there is a huge variety of options, also with high variability on the quality of projects (leading to widespread green-washing claims). 
\end{itemize}

The origin of carbon credits can be traced back to the Kyoto Protocol and the UNFCCC CDM (Clean Development Mechanism). The CDM defined dozens of methodologies to properly account for avoidance or removal of carbon, as each specific sector/application involves its own set of complexities. The complexity of Carbon Markets led to the formation of the Taskforce on Scaling Voluntary Carbon Markets (TSVCM) which in 2021 published one of the most comprehensive reports on the subject \cite{TSVCM2021}. TSVCM was later rebranded to the Integrity Council for the Voluntary Carbon Market (ICVCM) \cite{ICVCM2022}. ICVCM offers an independent governance body for the voluntary carbon market. Their purpose is to ensure the voluntary carbon market accelerates a just transition to 1.5C and it brings together the most important international organisations in this space. It is also important to mention the Voluntary Carbon Markets Integrity Initiative (VCMI) \cite{VCMI2022}. VCMI is focusing on corporate claims while ICVCM aims to address the quality of the underlying carbon offsets. Both, ICVCM \cite{ICVCM_VCMI} and VCMI \cite{VCMI_ICVCM} issued statements in summer 2023 to publicise their cooperation to operationalise a high-integrity market to accelerate global climate action. 

Bringing a consistent set of "principles" (e.g. standards), represents the priority of ICVCM. In July 2023, ICVCM made the public release of their Core Carbon Principles (CCPs) and Assessment Framework (AF) \cite{ICVCM_CCP}. The 10 "Core Carbon Principles" (CCP) are \cite{ICVCM_CCP}:

\begin{enumerate}
    \item \textbf{Governance: Effective governance.} The program needs to establish strong governance mechanisms that ensure transparency, accountability, ongoing improvement, and overall high standards.
    \item \textbf{Governance: Tracking.} The program must employ a registry to uniquely identify, record, and monitor mitigation activities and issued carbon credits, ensuring a secure and clear means of credit identification.
    \item \textbf{Governance: Transparency.} The program is required to furnish extensive and transparent details about all accredited mitigation activities. This information should be accessible to the public in digital form and be easily understandable to individuals without specialised expertise, facilitating the thorough examination of mitigation efforts.
    \item \textbf{Governance: Robust independent third-party validation and verification.} The program must establish program-level mandates for rigorous independent third-party validation and verification of mitigation activities.
    \item \textbf{Emissions Impact: Additionality.} The reduction/removal emissions resulting from the mitigation activity must be considered as additional, meaning that they would not have taken place without the incentive generated by the revenue from carbon credits.
    \item \textbf{Emissions Impact: Permanence.} The reductions/removals achieved through the mitigation activity must be either permanent or, in cases where there's a risk of reversal, the program should have measures in place to mitigate those risks and provide compensation for any reversals.
    \item \textbf{Emissions Impact: Robust quantification of emission reductions and removals.} The reductions/removals of emissions resulting from the mitigation activity must be accurately measured using rigorous and conservative methodologies that encompass completeness and scientific rigour.
    \item \textbf{Emissions Impact: No double counting.} The reductions/removals from the mitigation activity must not be counted twice; they should only be counted once to achieve the mitigation targets or goals. This prohibition on double counting includes avoiding scenarios like double issuance, double claiming, and double use.    
    \item \textbf{Sustainable Development: Sustainable development benefits and safeguards.} The carbon-crediting program shall have clear guidance, tools and compliance procedures to ensure mitigation activities conform with or go beyond widely established industry best practices on social and environmental safeguards while delivering positive sustainable development impacts.  
    \item \textbf{Sustainable Development: Contribution toward net zero transition.} The mitigation activity shall avoid locking-in levels of GHG emissions, technologies or carbon-intensive practices that are incompatible with the objective of achieving net zero GHG emissions by mid-century. 
\end{enumerate}

Most of the principles stated above are obvious and agreed among all carbon/standard registries (e.g. Gold Standard, Verra) but some can be complex to agree on. For example, Permanence is a problem for AFOLU projects (Agriculture, Forestry and Other Land Uses) but it's usually straightforward for Energy projects. Probably the most complex principle of all relates to “Additionality” as many organisations disagree on what exactly this can constitute for specific scenarios.

\subsection{DLTs/Blockchain}

\begin{figure}
  \includegraphics[width=\linewidth]{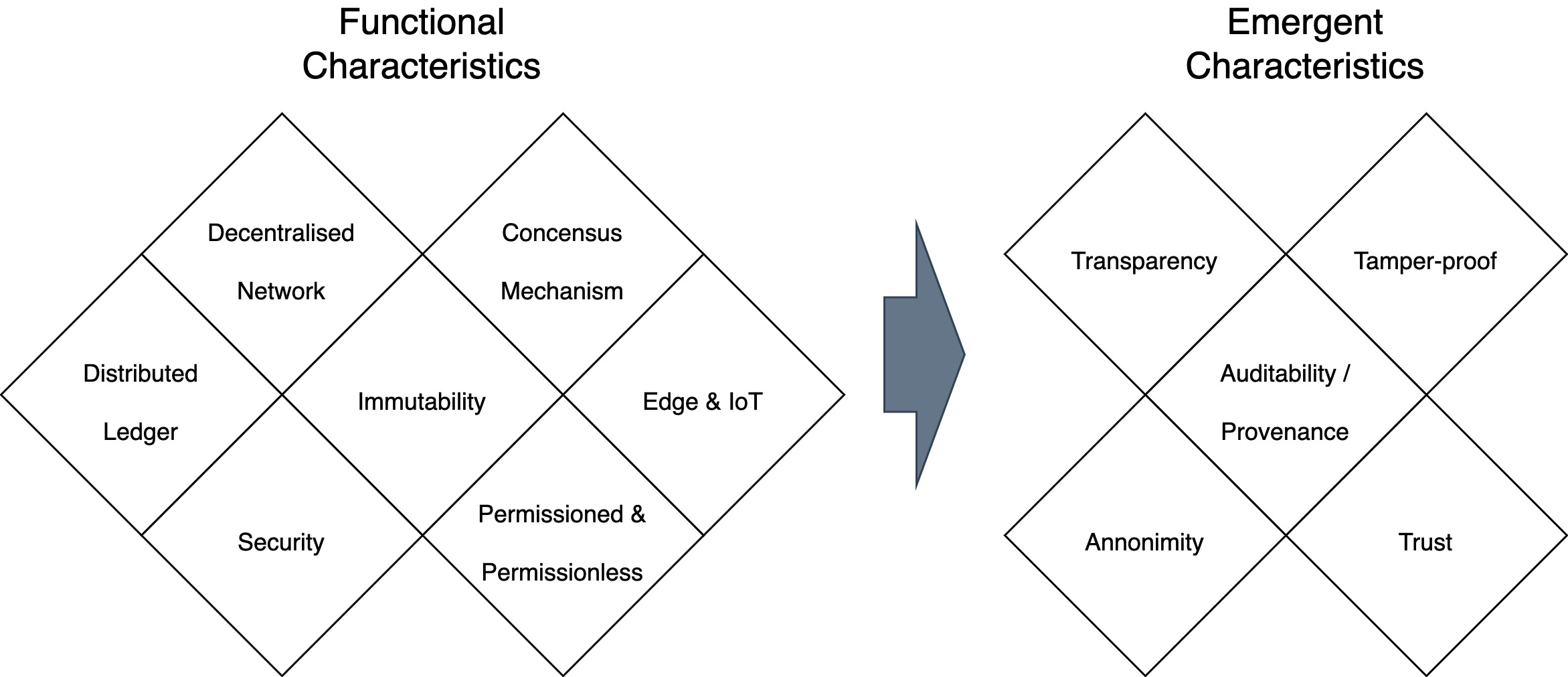}
  \caption{DLT/Blockchain Characteristics.}
  \label{fig:BlockchainOverview}
\end{figure}

Distributed Ledger Technology (DLTs) encompasses a broad range of databases with transactions that are shared and replicated across a network of computers. It allows multiple parties to record transactions and share the same version of the truth, without the need for a central authority. Blockchains are one of the many types of DLTs. DLTs use cryptography and consensus algorithms to ensure the integrity and security of the data. They are often decentralised, meaning that there is no single point of control, and the ledger is maintained by a network of participants rather than a single entity. Figure \ref{fig:BlockchainOverview} provides an overview of functional and emergent characteristics which help to describe DLTs and more importantly to highlight their benefits in most application domains (including carbon markets).  

There are many types of DLTs, depending on which characteristics we require for the specific application, for example 'permissionless vs permissioned', consensus mechanism (e.g. Proof of Work, Proof of Stake) among many others. In the specific case of Carbon Markets, the growing importance of “Governance” and “Regulations” means that well known financial requirements such as KYC (Know Your Customer) and AML (Anti Money Laundering) are becoming essential. 

Looking at the CCPs described in the previous section (e.g. real, MRV, counted once) it becomes obvious how the core requirements of a good functioning carbon market could benefit from DLT systems as they can provide a secure and transparent way to track and verify carbon credits. By using DLTs, it is possible to create a transparent and secure record of carbon credits that cannot be altered or tampered with. This can help to ensure the integrity of the carbon market and encourage its scalability. Moreover, DLT can be used to automate the buying and selling of carbon credits through the use of smart contracts. Smart contracts are self-executing contracts with the terms of the agreement between buyer and seller being directly written into lines of code. This can help to streamline the carbon credit trading process and reduce the need for intermediaries.

\subsection{Ecosystem}

As the above areas merge and mature, there are specific terms/concepts that are relevant to this Ecosystem. A great example of this comes from the Climate Warehouse \cite{ClimateWarehouse}, an initiative proposed by the World Bank that has been running for the past few years completing a number of pilots with Governments and many other market stakeholders. In recent years, all these efforts have converged to the Climate Action Data Trust (CAD Trust)\cite{CADT}. 

CAD Trust offers a decentralised, blockchain-powered digital infrastructure that connects these registries and provides public access to the information in a harmonised and user-friendly format. Their objective is to maximise the transparency of carbon credits, minimise the risk of double counting, and enhance the overall integrity of the markets. With a decentralised and secure digital infrastructure, the services built by the public and private sector such as compliance reporting, transacting, and benchmarking can be easier and more effective. Figure \ref{fig:EndToEndEcosystem} follows the pattern of the Climate Warehouse / CAD Trust \cite{Shiraishi2022, CADT} using the concepts of layers which helps to understand all the different components and their interaction. 

\section{Recent Developments (State of the Art)}

Probably the first published work on combining carbon credit and DLT comes from Leonhard \cite{Leonhard2017}. His proposal developed a solution that allowed teams of academics to issue carbon credits as crypto currency on the Ethereum network. Mok et al. \cite{Mok2019} in 2019 provides one of the first World Bank examples of testing the use of blockchain to build a meta-registry for decentralised climate markets. Franke et al. \cite{Franke2020} examines the benefits and constraints of applying blockchain technology for the Paris Agreement carbon market mechanism and develops a list of technical requirements and soft factors as selection criteria to test the feasibility of two different blockchain platforms.

Other applications of DLT in carbon/energy comes from Ashley et al. \cite{Ashley2018} which describes a solution for autonomous blockchain of custody for renewable energy credits and carbon credits that leaves a simple audit trail, significantly reducing the associated time and cost, and enabling producers to monetize their credits immediately after generation. Their work cited the Clean Energy Blockchain Network (CEBN) and other real-life experimental deployments. Pan et al. \cite{Pan2019} provides a similarity assessment between the mechanism of carbon trading and blockchain. The study confirms the potential value of the technology on this specific domain. Kim et al. \cite{Kim2020} proposes a blockchain-based carbon emission rights verification system to learn proven data further by using the governance system analysis and blockchain mainnet engine to solve these problems. Probably the most comprehensive and up-to-date reviews on the topic of TRLs (Technology Readiness Levels) for DLT + Carbon Markets comes from Sipthorpe et al. \cite{Sipthorpe2022}. The paper provides an objective, academic review of blockchain solutions for carbon markets. In surveying the entire market, they found the current ecosystem is diverse, fragmented, and relatively immature.

In terms of standards focused publications, one of the first works in the space of DLT standards comes from Anjum et al. \cite{Anjum2017}. Their work compares popular DLTs across key areas such as consistency, scalability, security principles and many other areas. The overview is aimed at practitioners looking to expand beyond the traditional finance space to all other areas of applications (e.g. supply chain, healthcare). Deshpande et al. \cite{Deshpande2017} provides a comprehensive overview of DLTs Challenges, opportunities and the prospects for standards. As a BSI (British Standards Institution) report, it provides perspectives across short, medium and long term standards on the growing DLT space. 

Hyland-Wood et al. \cite{Wood2018} argue about the need for standards in the DLT space by summarising some existing international standards  work  and propose directions for additional standards development that could meaningfully be explored in the near future without negatively impacting additional invention. Marsal-Llacuna et al. \cite{Marsal2017} toke a different angle on the topic of DLT standards by discussing new standard drafting models emerging from the disruptive blockchain community. These models can challenge traditional standard development bodies. Gramoli et al. \cite{Gramoli2018} listed standards organisations and the efforts they devote to standardise DLT. They identify a lack of terminology that can hamper communication on this topic and propose clarifications to address these ambiguities. Finally, they propose a high-level description of DLT by describing three elements of their functional architecture (consensus, security and ownership). Lima \cite{Lima2018} provides a case of the importance of DLT standards, providing one of the first classification of DLT/blockchain standards and a comprehensive overview of the IEEE ecosystem.

Aristidou et al. \cite{Aristidou2019} provides an assessment of DLT standards, currently and future, with a focus on government applications. This aims to consider applications that tackle corruption, fraud, lack of transparency, alienation and disconnection of citizens from decision-making centres to ensure proper governance conditions are offered to citizens. Bello et al. \cite{Bello2019} focused on comparing traditional payment systems standards within blockchain systems. For example, they compare the Payment Application Data Security Standards (PA-DSS) applicability towards transaction-supported blockchain platforms to test the standard's applicability. Drljevic et al. \cite{Drljevic2020} paper explores standards and risk as factors, which can support or hinder the sustained application of blockchain in a broad scope of environments. They conclude that a gap exists in normative frameworks that affect the adoption and sustainable use of blockchain technology. Closing this gap can support the sustainable use of blockchain technology. König et al. \cite{Konig2022} provides an overview of standardisation work about blockchains and DLTs, a set of comparison criteria for future work and a comparison of the existing standards work itself. With that information, aligning to existing standardisation efforts becomes easier, preventing duplication and supporting prioritisation (exactly the main argument of the present paper but across a broader domain). Naga et al. \cite{Naga2021} explains the standards and categorization of blockchain and its protocols. It includes the protocols of network and data organisation, distributed consensus protocols, the framework of autonomous organisation that relies on smart contracts executed in distributed virtual machines, and the human–machine interface implementations. Tan et al. \cite{Tang2021} reviews a total number of 167 blockchain standard projects. Highlighted topics including foundational, security, privacy \& identity, application, data and test \& assessment. Zhi et al. \cite{Zhi2021} summarised the current status of blockchain standardisation in China and abroad, with particular focus on the fields of energy and power. The paper also proposed a standardisation framework for these sectors.

The importance of standards in developing well functioning markets is undeniable. We can for example cite the work of Perrons et al. \cite{Perrons2020} which puts forward case study evidence from the Intel Corporation and the Energistics Consortium showing what the geoenergy sector can learn about blockchain from other industries, and highlights that the absence of data standards and interoperability has contributed to blockchain's failure to deliver significant value in the geoenergy domain thus far. Keogh \cite{Keogh2020} sets the case of how a standardised framework could enhance food traceability, drive FSC efficiencies,enable data interoperability, improve data governance practices, and set supply chain identification standards for products and assets (what), exchange parties (who), locations (where), business processes (why), and sequence (when). In more general terms, Schmidt et al. \cite{Schmidt2022} quantifies a novel channel that contributes to greater trade integration: the release of harmonised, voluntary product standards. Their study shows that harmonised standards have contributed up to 13\% of the growth in global trade. 

To the best knowledge of the author, this work represents the first overview of DLT and carbon markets focused on the role of "standards". This work is particularly unique in its comprehensive and up-to-date approach that covers the intersection of DLT and carbon markets as shown by Figure \ref{fig:EndToEndEcosystem}. 

\subsection{Challenges and Opportunities}

The growing area of DLT and Carbon Markets has experienced multiple challenges, particularly around credibility. As an example we can cite the debate between Verra and Crypto players like Toucan/KlimaDAO \cite{LedgerInsightsVerra, SPVerra}. The problem emerged because Toucan had tokenized millions of retired Verra-issued credits. Verra wants to make sure that retiring a carbon credit means that it has been fully used to offset a company's emissions and cannot be used again. However, the work done by Toucan allowed for a digital version of that carbon credit, called a BCT, to be traded as a digital commodity, which undermines the original intention of the carbon credit. 

\begin{figure}
  \includegraphics[width=\linewidth]{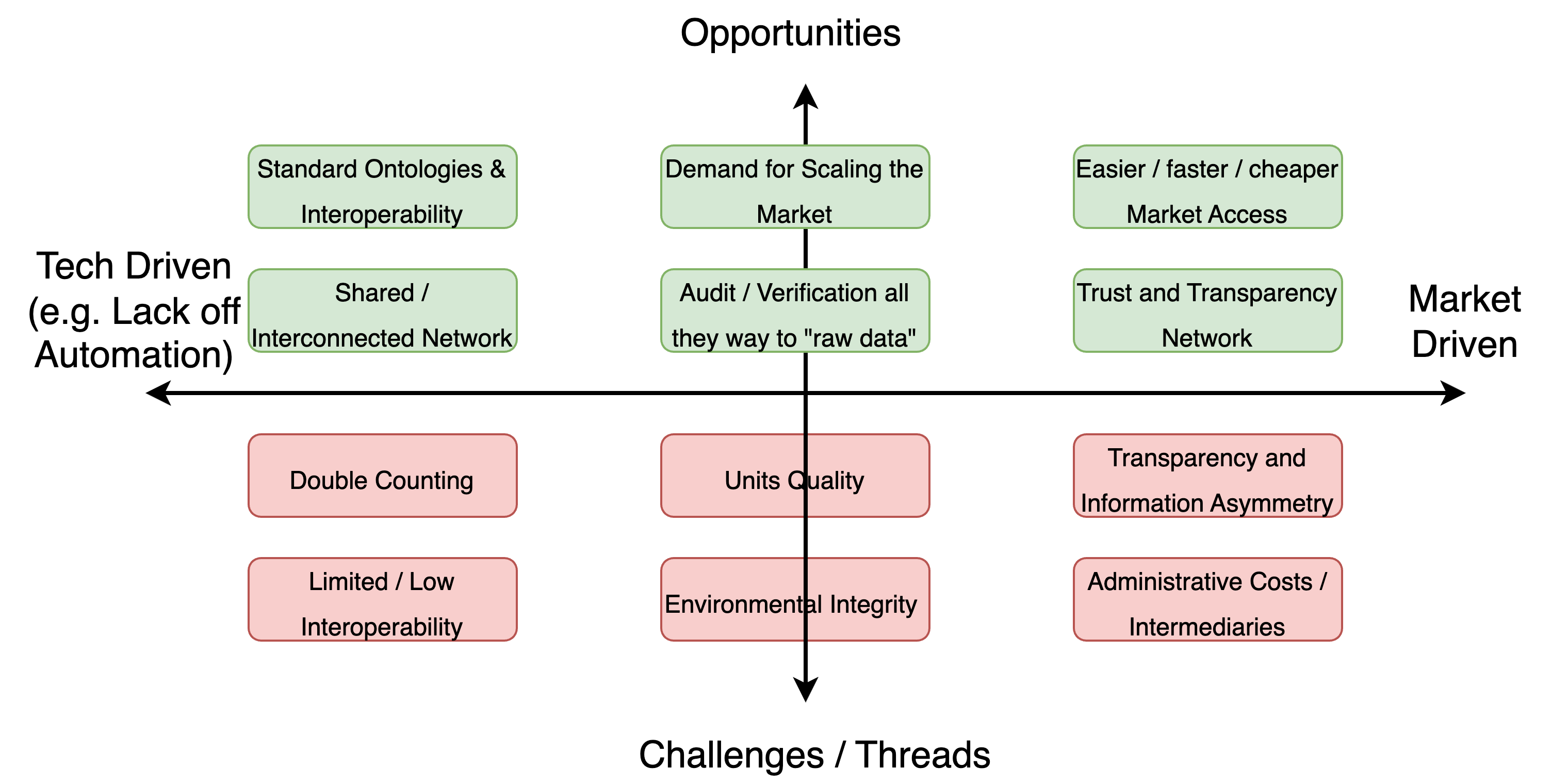}
  \caption{Challenges and Opportunities in the growing space of DLT + Carbon Markets.}
  \label{fig:ChallengeOpportunity}
\end{figure}

It is worth mentioning that Verra is not against the use of technology like DLT, but against the lack of honesty/transparency, e.g. \cite{Rackoff2022}. In August 2022 Verra launched a public consultation period during which it collected input from market participants and potential partners on how Web3 can be properly brought into the fold \cite{VerraConsultation}. Other carbon credit standard registries such as Gold Standard has similar views \cite{GoldStandardDLT}. Figure \ref{fig:ChallengeOpportunity} provides an overview of this fact, Challenges vs Opportunities are usually related and represent two sides of the same coin. A great example of this comes from the UNEP blog "Blockchain solutions to carbon market challenges"\cite{UNEP2019}.

Despite bad experiences over the past years, there are examples of good practices that help move the field forward. In this regard we can cite the example of EEX admitting the blockchain-based ClimateTrade to the European mandatory carbon market \cite{LedgerInsightsEEX}. EEX also partnered with another blockchain-based voluntary carbon market company, AirCarbon Exchange from Singapore. It is important to highlight that EEX (European Energy Exchange AG) is one of the main companies appointed as a common auction platform for the EU ETS Market by the European commission \cite{EEX2022} (regulated EU carbon market). These positive examples show that when organisations adopt best practices and cooperate, it is possible to deliver real-life DLT applications in carbon markets. In the next section, specific recommendations will be made to help address some of the existing challenges and take advantage of the ongoing opportunities. 

\section{Areas of Future Standards Development}

\begin{figure}
  \includegraphics[width=\linewidth]{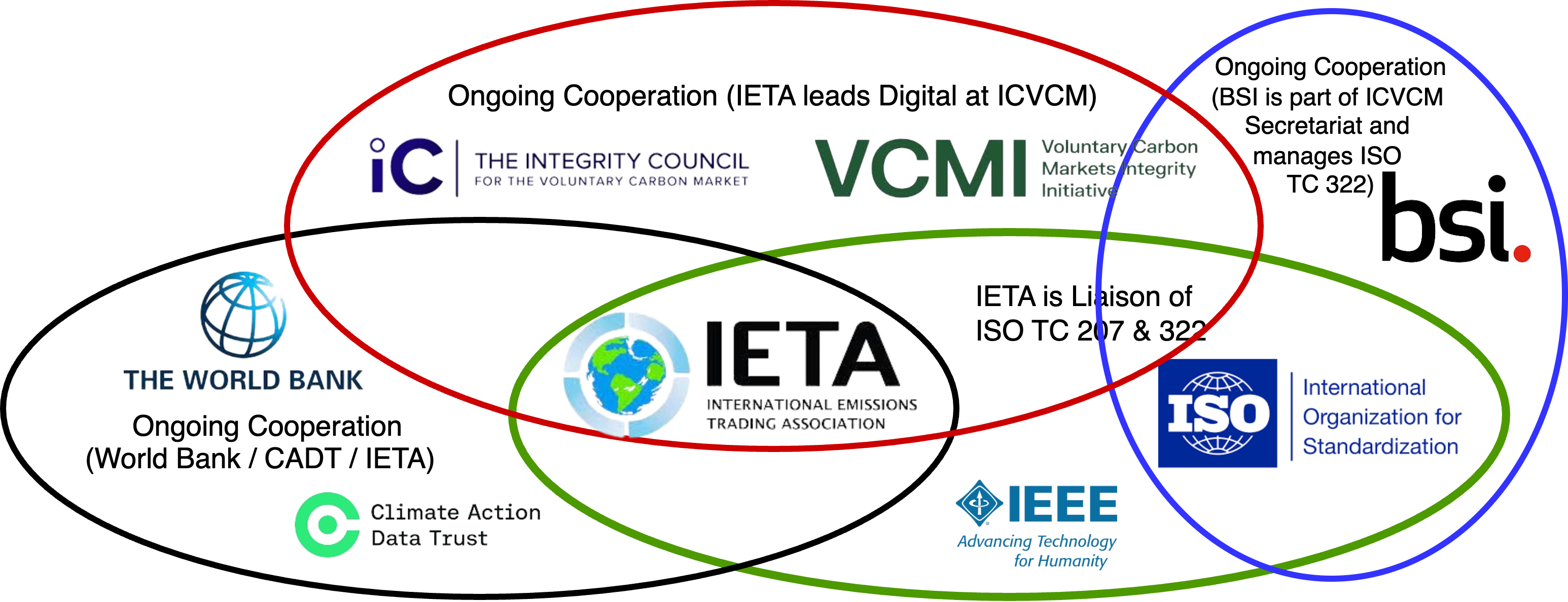}
  \caption{Overview of current non-profit international initiatives.}
  \label{fig:keyplayers}
\end{figure}

The future ecosystem of DLT and Carbon Markets should, broadly speaking, resemble that shown in Figure \ref{fig:EndToEndEcosystem}. In the following subsections each of these layers and components will be described, individually or as a group. In terms of key initiatives and non-profit organisations, Figure \ref{fig:keyplayers} provides a Venn diagram view. Having IETA in the centre demonstrates the focus of developing an ecosystem that is focused on market participants (IETA is the oldest and main international trading association for carbon markets).

\subsection{Markets}

The previous section should have demonstrated that Carbon Markets are extremely diverse and fragmented. Regulated markets are by nature fragmented and diverse because they follow national/regional priorities, for example alignment to the region's NDC (Nationally Determined Contributions) as part of the Paris Agreement. Fortunately, more cooperation is emerging on this front across countries, especially with the ongoing work in Article 6 of the Paris Agreement.

Voluntary Carbon Markets offer a less political avenue of progress. In this regard we should highlight the ongoing work that ICVCM and VCMI are trying to accomplish by trying to bring together the main Carbon Market stakeholders from around the world. This task involves less politics and more technical in-depth knowledge of what Core Carbon Principles (CCPs) should apply to all carbon credits.

In this front the main action point relates to engagement of international initiatives bringing multiple organisations together (including competitors) to address common challenges. The role of IETA (International Emissions Trading Association) should be highlighted as they are engaged with all the main international initiatives in this space, namely:
\begin{itemize}
    \item ICVCM. IETA is part of the governing body of the ICVCM and in fact they are in charge of the "digital assets" work stream (the ICVCM group focused on DLT and other digital technologies). 
    \item World Bank and UNFCCC. IETA continuously engages these organisations and more importantly contributed to the ongoing work taking place in CAD Trust.
\end{itemize}

In terms of specific types of "Standards" that are expected to be covered in this area, we can cite (but not be restricted) to the following examples:  
\begin{itemize}
    \item Contracts Types. IETA and other market makers have a long history of setting common international standard contracts for carbon credits. This is expected to continue under a digital (DLT) framework but with the added value of making contracts "smart" and therefore supporting a wide range of automations.  
    \item Project Types. ICVCM emerged primarily to help develop a universally accepted set of CCP (Core Carbon Principles). They are setting and enforcing a definitive global threshold, drawing on the best science and expertise available, so high-quality carbon credits efficiently mobilise finance towards urgent mitigation and climate resilient development. There is still extensive ongoing work on this space across many different market participants for specific methodologies. 
\end{itemize}

The above examples are essential to develop mature and liquid international carbon markets. In fact, clarifying project types is essential in order to know if fungible or non-fungible tokens (NFTs) are to be created. This is still an active area of discussion but most likely DLT carbon markets will require a range of "fungible NFTs". \footnote{Fungible NFTs refers to the idea that adding/splitting emissions from the same project type (same CCPs) is ok (fungible tokens) while between different project types is not (NFTs).}

In Figure \ref{fig:EndToEndEcosystem} market standards fall mainly into the "Registry", "Metadata" and "Service" Layers. ICVCM CCPs is a perfect example for the registry layer while the Data Model from the Climate Warehouse / CAD Trust is a good example for the metadata layer\cite{CADT}. In terms of the service layer, there are multiple standards that can be applicable, for example Compliance Reporting could leverage International Standards such a GRI or the new IFRS ISSB (depending strongly on the specific service/application). ISO Technical Committees such as TC 68 (Finance) and TC 322 (Sustainable Finance) also have a key role to play in this space such as for example GLEIF / ISO 17442 (Legal Entity Identifier) or ISO 6166 (International Securities Identification Number - ISIN).

\subsection{Governance \& Regulations}

Governance is an essential area for any well functioning organisation. Unfortunately, few scandals in the crypto space in recent years can ultimately be traced back to poor governance (e.g. lack of processes, controls, checks). For example, the fall of FTX can be traced back to the liquidity crisis of the FTT token and his trading firm, Alameda Research Company, as Reuters reported that Bankman-Fried moved up to \$10 billion in FTX customer funds to Alameda, whose assets were primarily held in the FTT token \cite{FTXCollapse}. Despite all the hype and work in areas such as DAOs (Decentralised Autonomous Organizations), DLT projects and organisations can still exhibit poor governance practices. Recent studies demonstrate that many popular DAOs are actually extremely centralised (most tokens are concentrated in the hands of few accounts/individuals)\cite{Chainalysis}. 

One of the main reasons why Governance can be poor among DLT projects is because some support the ethos of DeFi (e.g. anyone in the world can lend money to anyone else, without the intervention of incumbents such as government or traditional banks). This ethos can contradict proper regulatory oversight which means that many of the great characteristics of DLTs could be compromised, e.g. Trust or Transparency can be jeopardised by poor KYC/AML.

The specific types of standards that are expected to be covered in this area include: 
\begin{itemize}
    \item CCPs. This focus area of standards aims to bring integrity and minimum levels of quality of carbon credits. ICVCM represents the most recent international initiative in this space, in fact ICVCM is an independent "governance body" for the voluntary carbon market (bringing together many other non-profit carbon credit standard bodies like Verra or Gold Standard). 
    \item Organisation Governance. The growing importance of "Sustainability" and "ESG" (Environment, Social and Governance) considerations across the corporate and financial world has given rise to many initiatives in this space. Some examples include: ISO 37000:2021 series and B Corporation Certification \cite{BCorp}. 
    \item Regulators Involvement. This refers specifically to VCM (Voluntary Carbon Markets) as regulated is well defined by each specific jurisdiction. 
\end{itemize}

As an example of regulators' involvement in VCMs we could cite the recent report and consultation from IOSCO\cite{IOSCO2022}. The Board of the International Organization of Securities Commissions (IOSCO) published in November 2022 a discussion report with the aim of advancing the discussion about what sound and efficient Voluntary Carbon Markets (VCMs) should look like and what role financial regulators may play in promoting integrity in those markets. As an international organisation, IOSCO is in close contact with important national regulators like the SEC in the US and the FCA in the UK. In Figure \ref{fig:EndToEndEcosystem} governance standards or regulators involvement covers everything, particularly aspects not shown in such figure and which relate to the organisations involved in the ecosystem. 

\subsection{Physical Layer}

In Figure \ref{fig:EndToEndEcosystem} the physical layer can relate to a wide range of options, each dependent on the specific economic sector. Although all economic sectors emit emissions, there are four which are particularly important due to their large proportion of emissions generation: 1) Energy, 2) Buildings and Materials (this also involves Heavy Industries such as Cement, Steel), 3) Transport and 4) AFOLU (Agriculture, Forestry and Other Land Uses). 

Standard areas that are of importance in the physical domain include: 
\begin{itemize}
    \item Carbon Accounting (CO2). The most important standard initiatives in this space are the GHG Protocol\cite{GHGP} and the ISO 1406X Series (developed by ISO TC 207). ISO and GHGP are actually extremely similar and therefore either could be used/extended. The main advantage of GHGP is that it is openly available, while the advantage of ISO is that it provides a comprehensive set of standards for verification and accreditation (ISO 14065 and 14066).  
    \item Energy. In most sectors, emissions and energy are directly linked. ISO 50000 standard demonstrates this very clearly, but the fact is that different economic sectors can offer more relevant domain specific options. For example, in the built environment BREEAM\cite{BREEAM} represents a top international option.  
    \item Other GHG (Greenhouse Gas) Emissions. CO2 is actually not the worst man-made emission, for example methane (CH4) common in agriculture is 25x more damaging than CO2, while nitrous oxide (N2O) is 298x\cite{CO2e}. Carbon credits are normally issued in CO2e (CO2 equivalent) metrics which aims to account for this fact.  
    \item Hardware \& Software. At the physical layer, in addition to "carbon" or "domain specific" standards we should add the role that sensor network standards could play. Ensuring that a reading/measurement actually represents the physical world encompasses a wide range of technologies. Standards on "Blockchain and IoT Integration" represent a good example of this broad and expanding space.
\end{itemize}

Although general methodologies to measure GHG are now well established with internationally recognised standards (e.g. GHGP, ISO 1406X), there is still ongoing R\&D (and future Standards) to properly assess emissions and baselines for specific economic sectors. One huge area of concern and ongoing work is represented by Scope 3. \footnote{Scope 3 emissions, also known as indirect emissions, are emissions that result from activities that are not directly controlled by an organisation but are related to its operations. These emissions can occur along the value chain, including upstream and downstream of an organisation's operations.}

As an example of this ongoing work, we can cite the recent work in the Transport sector. Organisations such as the Smart Freight Centre has produced the GLEC Framework \cite{SFC2022} which provides a specific standard for "Quantification and reporting of greenhouse gas emissions arising from transport chain operations" (ISO 14083 published in 2023). GLEC / ISO 14083 builds on top of generic standards such as GHGP or ISO 1406X to provide more accurate emissions quantification. This can hugely influence CCPs such as MRV, Real and Baselines. 

\subsection{Database Layer}

This layer of the ecosystem relies mainly on very generic standards and regulations (not specific to Carbon Markets or DLT). All data driven businesses, specially those that can gather data from a wide range of sources (e.g. IoT, Satellite) require a multi-stage data processing pipeline, from raw data entries (e.g. sensor, image) to valuable insights (carbon units). The main broad areas of work under this layer include:

\begin{itemize}
    \item Cyber Security. Any data platform should be protected against cyber attacks, especially when the system contains valuable datasets. Well-known standards in this space include certifications from ISO 27000 and SOC 2. 
    \item Privacy and Confidentiality. Ultimately, all emissions data relates to physical systems that relate to people or organisations. Due to a growing set of regulations involving personal information, it should be properly handled (anonymously). Additionally, organisations could have issues with confidentiality (sharing data between competitors requires careful considerations). 
\end{itemize}

The complexity in this layer of the ecosystem is that it needs to establish a delicate balance between accurate quantification of emissions (essential for meeting several CCPs) and privacy or confidentiality concerns. Audit \& verification of data and its processes is also essential. Getting this balance right could lead to amazingly fair and accurate "baselines" (e.g. baseline transport emissions for a disabled person can not be the same as that of an able adult).

\subsection{Transaction Layer}

Similar to the Database layer, the transaction layer involves aspects that are common for those working on financial transactions. A recent article by CoinDesk called "2023: The Year of Regulation vs. Decentralization" \cite{CoinDesk2022} demonstrates how recent catastrophic failures in the crypto space is leading to regulators like the SEC and CFTC to step in the coming years. Some specific areas of work on this layer include: 

\begin{itemize}
    \item Identity (e.g. KYC \& AML). International organisations like the Financial Action Task Force (FATF) have a long track record in setting standards/guidance in this space. One recently release standard that has substantially grown in importance among the financial sectors in the EU is ISO 17442 \footnote{The Global Legal Entity Identifier (LEI) System is designed to uniquely and unambiguously identify participants in financial transactions \cite{GLEIF2022}}. 
    \item Payments. In this domain we should highlight the role of ISO 20022. A single standardisation approach (methodology, process, repository) to be used by all stakeholders of international payments. Examples of users include important international organisations like the Swift network \cite{Swift2022}.
    \item Interoperability. This is probably the most needed, but also the most complex aspect of the DLT ecosystem. Many organisations, national and international, could be cited regarding ongoing work. The above literature review covers some good papers that touch on this point. A reasonable landscape of initiatives is provided by the EU\cite{EUBlockcchain2022}. We should specially highlight the role of broad international initiatives like ISO TC307 (Blockchain and distributed ledger technologies) and TC68/TC322 (Finance and ESG).
\end{itemize}

Each of the listed areas above can encompass a large number of standards. For example, although LEI (ISO 17442) has proven amazingly successful for legal entities / organisations, we still don't have something similar for individuals (unfortunately this is a more challenging task due to many privacy concerns). 

In this layer we should also highlight the recently published "Draft Standard for Using Blockchain for Carbon Trading Applications - IEEE 3218" \cite{IEEE3218}. This standard specifies technical framework, application processes and technical requirements for carbon trading applications based on blockchain, including functions, access, interface, security, and carbon consumption voucher coding. Currently, IEEE C/BDL (group responsible for IEEE 3218) and ISO TC322 experts are engaged to continue mapping and drafting relevant standards in this broad space, see Figure \ref{fig:keyplayers}.

\subsection{Other Digital Elements}

There are many other digital elements that require standardisation to achieve a fully functioning ecosystem, especially elements that help connect layers in Figure \ref{fig:EndToEndEcosystem}. Some specific areas include:  

\begin{itemize}
    \item Wallets. DLT Wallets are digital accounts that enable users to manage their crypto holdings. Some of these are specific to L1 DLTs and extensive work is underway across many domains to support international standards (e.g. ISO TC307). 
    \item Tokens. As described previously, tokenization represents an important area of work as we may require fungible and non-fungible (NFTs) tokens for supporting a comprehensive international carbon market.  
    \item Smart Contracts. These lines of code that automate transactions can be directly related to traditional carbon credits contracts, specially to support guarantees for important CCP such as permanence or insurance claims. For example, if a fire/flood destroys a forest, carbon credit investors are automatically compensated.
\end{itemize}

Probably the biggest challenge in this space comes from the development of standards that work across different Layer 1 DLTs. For example, ERC 20 is a very common standard in Ethereum but having common standards across all existing DLTs is a much more challenging endeavour due in part to the different characteristics of different DLTs (not just due to consensus mechanisms or protocols). All this eventually falls back into the broad "interoperability" umbrella. 

\section{Conclusion}

The present document has provided a comprehensive up-to-date overview of the DLT for Carbon Market space. Figure \ref{fig:EndToEndEcosystem} layered approach aims to break a very complex and fast changing ecosystem. As mentioned in the introduction, the paper aims to address two key challenges that are hampering progress, namely: Lack of coordination between key international organisations and lack of clarity on what this domain involves. The second point is now hopefully covered while the first one requires an ongoing approach between all the stakeholders engaged in this growing space. Some specific recommendations with which this paper can conclude are:

\begin{itemize}
    \item DLTs for the Carbon Markets ecosystem would benefit from a commonly used breakdown of components as shown in Figure \ref{fig:EndToEndEcosystem}. A holistic view of the ecosystem is important to identify priorities \& gaps that are hindering the ultimate goal, while a focused approach is important to deliver successful solutions (carbon market experts do not need to learn the details of DLTs but they need to understand how all fits together, and vice-versa).
    \item R\&D is still extremely important across the ecosystem (specially when involving new tech). As discussed across the present document, some areas/layers have well established concepts and standards but others are still in their infancy. Even topics such as carbon measurement with well-known standards such as GHGP or ISO 1406X are being extended to accommodate recent developments and further accuracy.
    \item Engagement across all stakeholders is key for progress. The great experience gathered from the Climate Warehouse / CAD Trust demonstrates the huge progress that can be achieved when working together, including between competitors. Although some for-profit players aim to move fast (typical of tech startups), it is unlikely any single for-profit organisation will completely dominate the market. This is not just due to the size/fragmentation of the market but also to the growing role of regulators.  
    \item In order to prevent the GIGO concept continuing causing havoc on the DLT + Carbon Market space, it is essential to understand that the ecosystem is only as strong as its weakest link. For example, we could develop amazing standards for DLT interoperability, but if carbon credit registry quality is poor, the ultimate output will be poor (Garbage). 
\end{itemize}

Standards are known to significantly influence trade, economic growth, and opportunities for innovation. It is the view of the author that a lack of minimum standards (best practices) is hindering the entire ecosystem of DLTs for Carbon Markets. As this paper describes, the challenge is to bring together in a comprehensive way a wide range of layers which are evolving fast, each at their own pace and under its own requirements. Hopefully the present work helps to bring anyone to the same level of understanding in order to address the gaps/priorities areas. 

\subsection{How to get involved?}

An important goal of the present work is to encourage cooperation and prevent fragmentation of a very young and growing market. The ongoing climate crisis requires scalable solutions that deliver ground impact fast (high integrity solutions). All the non-profit organisations listed in here are approachable using a variety of means. Either via memberships models for companies (e.g. IETA is an industrial association) or via national standardisation bodies (e.g. most national standards organisations are part of ISO). 

The suggested path to engagement in order of priority involves:
\begin{itemize}
  \item Governments: Depending on which branch of government, these could join many of the conversations, specially when connecting via UNFCCC (e.g. Paris Agreement, COP). 
  \item Non-profit: This can involve think tanks, charities, research, universities and many other types of non-profit organisations interested in supporting/engaging this space. Any of these could engage ISO via their national bodies. 
  \item For-profit: Companies such as startups or incumbents are better placed joining industrial bodies, in this regard we should highlight the role of IETA. Due to conflict of interest, for profit organisations can not always directly engage government or organisations like ISO (unless otherwise specified). In these circumstances it is better to engage the organisations mentioned here as individuals if possible. 
  \item Individuals: Many organisations mentioned here (e.g. ISO, IEEE, ICVCM) can be engaged with on a personal basis. For example, some national standards bodies (ISO) can allow for "experts" (e.g. independent consultant) to join committees on a personal basis (not necessarily linked to a specific organisation - although this depends on national standards procedures).  
\end{itemize}




\section{Acknowledgement}

The author would like to acknowledge the non-financial support of non-profit organisations and colleagues at ISO TC322, ISO TC307, ISO TC207, IEEE (specially C/BDL/BCN), IETA (specially Alasdair Were) and Imperial College London (specially Prof. Julie McCann). Regular meetings and valuable updates from their members helped to provide a more comprehensive view of the growing field of "Blockchain/DLTs for Carbon Markets". 



\section{Conflict of Interest}

P.M. Baiz is co-founder and CTO of Clearly, a startup helping Corporate and Financial Institutions transition to net zero transportation more efficiently (Higher RoI, Lower Risks, Higher Impact). He is also an Honorary Senior Research Fellow at Imperial College London.



\appendix



\bibliographystyle{IEEEtran}

\bibliography{export.bib}



\end{document}